\documentclass[pra,twocolumn,groupedaddress,showpacs,floatfix]{revtex4}

\usepackage{graphics}
\usepackage{graphicx}
\usepackage{bm}
\usepackage{amsmath}
\usepackage{amsfonts} 
\usepackage{amssymb}
\usepackage{latexsym}
\usepackage{color}
\usepackage{ctable}

\begin{document}

\title{Improved Eavesdropping Detection in Quantum Key Distribution}
\author{Muhammad~Mubashir~Khan,$^1$ Jie~Xu,$^1$ and Almut~Beige$\,^2$}
\affiliation{$^1$The School of Computing, University of Leeds, Leeds LS2 9JT, UK \\
$^2$The School of Physics and Astronomy, University of Leeds, Leeds LS2 9JT, UK}

\date{\today}

\begin{abstract}
Employing the fundamental laws of quantum physics, Quantum Key Distribution (QKD) promises the unconditionally secure distribution of cryptographic keys. However, in practical realisations, a QKD protocol is only secure, when the quantum bit error rate introduced by an eavesdropper unavoidably exceeds the system error rate. This condition guarantees that an eavesdropper cannot disguise his presence by simply replacing the original transmission line with a less faulty one. Unfortunately, this condition also limits the possible distance between the communicating parties,  Alice and Bob, to a few hundred kilometers. To overcome this problem, we design a QKD protocol which allows Alice and Bob to distinguish system errors from eavesdropping errors. If they are able to identify the origin of their errors, they can detect eavesdropping even when the system error rate exceeds the eavesdropping error rate. To achieve this, the proposed protocol employs an alternative encoding of information in two-dimensional photon states. Errors manifest themselves as quantum bit and as index transmission errors with a distinct correlation between them in case of intercept-resend eavesdropping. As a result, Alice and Bob can tolerate lower eavesdropping and higher system error rates without compromising their privacy. 
\end{abstract}

\maketitle

\section{Introduction} \label{intro}

Based upon the fundamental laws of quantum mechanics, quantum cryptography provides a technique for sharing secret cryptographic keys with unconditional security \cite{grtz2002}. The underlying idea of quantum cryptography -- also known as quantum key distribution (QKD) -- is that the secret bits of the cryptographic key are encoded in the quantum states of photons. When the encoded photons are transmitted from one party to another, let us call them \emph{Alice} and \emph{Bob}, an eavesdropper, \emph{Evan}, cannot view the key without introducing a significant error rate. Measuring the bit transmission error rate by comparing some randomly selected test bits hence allows Alice and Bob to verify the security of their key transmission. Since the first protocol for quantum cryptography, BB84 \cite{BB84}, many variations of it have been proposed. These aim at increasing the eavesdropping error rate, flexibility, and efficiency of QKD. For example, Bennett \cite{B92} suggested in 1992 to use two non-orthogonal two-dimensional photon states to distribute keys, thereby increasing the flexibility in the choice of states. Later on, work in \cite{DB1998} and \cite{BG1999} introduced respectively a three-state scheme and a six-state protocol which create more difficulties for Evan.

The main difference between the protocol by Beige {\em et al.} \cite{Beige22002} and earlier protocols lies in the encoding of information into quantum states. Instead of encoding a whole alphabet into each basis, Ref.~\cite{Beige22002} proposes a scheme where all the vectors of the same basis encode the same letter. When optimising this strategy, a direct quantum communication protocol between Alice and Bob is obtained  \cite{Beige12002}. This protocol is a QKD protocol with maximum efficiency but a relatively low error rate in case of eavesdropping. Recently, Khan {\em et al.}~\cite{KMB2009} used the same encoding to maximise intercept-resend eavesdropping error rates. In the following we refer to this protocol as KMB09. When using two-dimensional photon states, KMB09 is essentially the same as the SARG04 protocol by Scarani {\em et al.}~\cite{hugo2}  with $\chi = 1/\sqrt{2}$ \footnote{This protocol is tailored to be robust against photon number splitting attacks. In SARG04, Alice publicly announces which one of the {\em four} different sets of states she used, while she uses only {\em two} sets in KMB09. This difference is due to some redundancy of the SARG04 protocol.}. Brierley~\cite{BS2009} increased the eavesdropping error rate of these protocols even further by allowing Alice and Bob to use multiple bases of more than two-dimensional photon states. This improvement comes at the cost of very low key bit transmission efficiencies.

In this paper, we design a QKD protocol with two-dimensional photon states which is characterised not only by one but \emph{two} different types of errors with a distinct correlation between both of them in case of eavesdropping. More concretely, we propose a QKD protocol, where Alice and Bob can calculate a Quantum Bit Error Rate (QBER). In addition, they can exchange information about the index of the transmitted state. Unexpected differences in the index are measured by the Index Transmission Error Rate (ITER). This combination of measures makes it much harder for Evan to escape his detection. The eavesdropper not only has to minimise different types of errors. Distinct correlations between the ITER and the QBER, ie.~a  distinct signature of eavesdropping, allows Alice and Bob to detect the origin of their errors. As a result, Alice and Bob can tolerate lower eavesdropping and higher system error rates without compromising their privacy. We no longer require that the error rates in case of eavesdropping unavoidably exceed the system error rate, since Evan can no longer disguise his presence as system errors. This should allow Alice and Bob to communicate securely over much longer distances.

The remainder of this paper is structured as follows. Section~\ref{KMB} analyses the KMB09 protocol~\cite{KMB2009} and derives  the error rates and key bit transmission efficiencies of this protocol for the case, where Alice and Bob use two-dimensional photon states. Although this has not been emphasized in our earlier work, the KMB09 protocol is a QKD protocol with a QBER and an ITER. However, as we shall see below, there is no distinct correlation between these errors in case of eavesdropping. Section~\ref{More} therefore presents a variation of the KMB09 protocol which permits Alice and Bob to use not only two but three different bases. The third basis increases the flexibility of the proposed protocol which can be used to create a {\em linear} dependence between the ITER and the QBER in the case of intercept-resend eavesdropping. Our calculations are illustrated by figures which take all possible measurement bases of Evan into account and show typical values for actual realisations of the proposed protocol. We finally summarise our findings in Section~\ref{Conclusions}. 

\section{The KMB09 protocol} \label{KMB}

Our first step in designing a QKD protocol with the above described features is to notice that the encoding of information used in the KMB09 protocol already results in two different types of errors: a QBER and an ITER. In our earlier work \cite{KMB2009}, we did not recognise the potential advantage of this fact and considered only the ITER when designing optimal QKD protocols based on more-than-two dimensional photon states. In this section we analyse the KMB09 protocol with two-dimensional photon states in much more detail than previously done and present analytical results for the QBER, the ITER, and the key bit transmission efficiencies with and without eavesdropping and introduce the notation used throughout this paper. As we shall see below, in the KMB09 protocol, there are only very weak correlations between the QBER and the ITER. The second step in designing a QKD protocol with an improved eavesdropping detection is therefore the introduction of a variation of the original KMB09 protocol in Section \ref{More}.  

\subsection{The basic protocol}

In the case of two-dimensional photon states, the KMB09 protocol suggests that Alice and Bob use two bases $e$ and $f$ of the form 
\begin{eqnarray}\label{eq:basisStatesGeneralKMB}
	e &\equiv & \left\{ |e_i \rangle: i=1,2 \right\} \, , \nonumber \\ 
	f &\equiv & \left\{ |f_i \rangle: i=1,2 \right\} \, .
\end{eqnarray}
Alice randomly prepares photons in the basis states of $e$ and $f$ and sends them afterwards to Bob. For each incoming photon, Bob randomly selects a basis and measures its state either in \textit{e} or in \textit{f}. After the transmission of many photons, Alice publicly announces the index \textit{i} of each prepared state. Bob compares these indices with the indices of his measurement outcomes and interprets his findings as suggested by Table~\ref{tab:states_interpretation1}. For each measured photon, he obtains either 0, 1, or $\times$. These results indicate respectively whether a $0$, a $1$ or \emph{no bit} has been transmitted. Under ideal conditions, it is not possible that Alice and Bob have different indices, when both use the same basis. A key bit is hence obtained whenever the index of the state sent by Alice is different from the index measured by Bob. Notice that this encoding and decoding works without having to apply any constraint to the basis states of $e$ and $f$. Bob obtains a 1, if he measures an $e$ state and he obtains a 0, if he measures $f$, whenever a key bit is transmitted, i.e.~when Alice's and Bob's indices are different. 

\begin{table}[t]
\begin{center}
\begin{tabular}{@{}|c|c|c|c|c|c|}
\hline
   & \multicolumn{4}{|c|}{Bob measures} \\ \cline{2-5}
		{Basis used}	& \multicolumn{2}{|c|}{Same index} & \multicolumn{2}{|c|}{Different index} \\
		\cline{2-5}
		{by Alice} 	& \hspace{0.1cm} \textsl{e} \hspace{0.1cm} & \hspace{0.1cm} \textsl{f} \hspace{0.1cm} & \textsl{e} & \textsl{f} \\
\hline
   $ e $ meaning 0 & $  \!\!\times \!\! $ & $ \!\!\times \!\! $ & $ {\rm error} $ & $0$ \\
\hline
   $ f $ meaning 1 & $  \!\!\times \!\! $ & $ \!\!\times \!\! $ & $  1 $ & $ {\rm error} $ \\
\hline 
  \end{tabular}
  \end{center}
\caption{Bob's interpretation of the measurements which he performs on the incoming photons, after Alice announced the index of the transmitted state. No key bit is obtained when both have the same index, as indicated by the crosses. An error occurs when both their states have a different index although both use the same bases.}
  \label{tab:states_interpretation1}
\end{table}

Suppose the secure key distribution between Alice and Bob is challenged by an eavesdropper \textit{\rm Evan}. As in Ref.~\cite{KMB2009}, we assume in the following that Evan applies an intercept-resend strategy in order to obtain the information transmitted from Alice to Bob. This means, Evan measures the state of each incoming photon in a basis $g$, 
\begin{eqnarray}\label{eq:basisStatesGeneralKMB2}
	g &\equiv & \left\{ \left| g_{k}  \right\rangle: k=1,2 \right\} 
\end{eqnarray}
which is optimal for his purpose of minimising the error rate introduced into the key bit transmission. Later on Evan forwards the photon, in the state he found it in, to Bob. 

\subsection{Parametrisation of states} \label{Bloch}

Restricting ourselves to two-dimensional photon states allows us to represent the basis states of $e$, $f$, and $g$ with the help of a Bloch sphere \cite{Knight}. The Bloch sphere representation allows to visualise two-dimensional complex vectors by writing them as three-dimensional real vectors. This is possible after neglecting an overall phase factor with no physical consequences. Taking this into account, any quantum mechanical state vector can be expressed as a function of two angles $\theta$ and $\phi$ which serve as the spatial coordinates for the corresponding three-dimensional real vectors of unit length. In the following we use this approach to obtain a relatively simple parametrisation for all the possible basis states which can be used by Alice, Bob, and Evan.

Without restrictions we assume in the following that the basis states of $e$ are the canonical basis vectors. Choosing an appropriate coordinate system, the vectors of the $f$ basis can then be written such that they relate to the states of $e$ via a simple rotation by an angle $\theta_1$. In this coordinate system, the generation of the basis states of $g$ requires in general two rotations of the $e$ basis: one rotation by an angle $\theta_3$ and another rotation by an angle $\phi_3$. More concretely, we assume in the following that   
\begin{eqnarray} \label{eq:general_efh_KMB}  
&& \hspace*{-0.7cm} \left| e_{1} \right\rangle = \left( \begin{array}{r} 1 \\ 0 \end{array} \right) \, , ~~
\left| e_{2} \right\rangle  = \left( \begin{array}{r} 0 \\ 1 \end{array} \right) \, ,
\nonumber \\
&& \hspace*{-0.7cm} \left| f_{1} \right\rangle = \left( \begin{array}{r}  \cos \left( \frac{1}{2} \theta_1 \right) \\  \sin \left( \frac{1}{2} \theta_1 \right)  \end{array} \right) \, , ~~
\left| f_{2} \right\rangle  = \left( \begin{array}{r}  \sin \left( \frac{1}{2} \theta_1 \right) \\  -\cos \left( \frac{1}{2} \theta_1 \right)  \end{array} \right) \, ,
\nonumber \\
&& \hspace*{-0.7cm} \left| g_{1} \right\rangle = \left( \begin{array}{r}  \cos \left( \frac{1}{2} \theta_3 \right) \\  {\rm e}^{{\rm i}\phi_3} \sin \left( \frac{1}{2} \theta_3 \right) \end{array} \right) \, , ~~
\left| g_{2} \right\rangle  = \left( \begin{array}{r}  \sin \left( \frac{1}{2} \theta_3 \right) \\ - {\rm e}^{{\rm i} \phi_3} \cos \left( \frac{1}{2} \theta_3 \right) \end{array} \right) \, , \nonumber \\
\end{eqnarray}
where $\theta_1$, $\theta_3$, and $\phi_3$ are real numbers ranging from $0$ to $2\pi$. These three parameters describe all possible sets of bases $e$, $f$, and $g$. When looking for an optimal strategy, Alice and Bob only need to consider all the possible values of these three angles.

Since all the states of the bases $e$, $f$, and $g$ are of unit length and pairwise orthogonal, one can easily check that 
\begin{eqnarray}\label{eq:basisStatesEveKMB}
\sum_{j = 1}^2 | \langle g_k | e_{j} \rangle |^2 \, = \, 
\sum_{j = 1}^2 | \langle g_k | f_{j} \rangle |^2 \, = \, 1 \,.
\end{eqnarray} 
This means, for example, that
\begin{eqnarray}\label{eq:basisStatesEveKMB2}
| \langle g_1 | e_2 \rangle |^2 = 1 - | \langle g_1 | e_1 \rangle |^2 = 1 - | \langle g_2 | e_2 \rangle |^2 
\end{eqnarray} 
which implies  
\begin{eqnarray} \label{eq:basisStatesEveKMB3}
| \langle g_1 | e_1 \rangle |^2 = | \langle g_2 | e_2 \rangle |^2 \, .
\end{eqnarray} 
An analog relation applies between the states of the $g$ and of the $f$ basis.

\subsection{Index transmission errors}

Let us first examine the error in the transmission of each state's index. An index transmission error occurs when a photon prepared in $|e_i \rangle$ (or $|f_i \rangle$) is measured at Bob's end as $|e_j \rangle$ (or $|f_j \rangle$ respectively) with $i \neq j$. Assuming that Alice randomly prepares all the states of $e$ and $f$ with the same frequency and that Bob measures $e$ and $f$ with the same frequency, we find that the Index Transmission Error Rate (ITER) of the KMB09 protocol equals~\cite{KMB2009}
\begin{eqnarray}\label{eq:P_ITER_KMB}
P_{\rm ITER} = 1 - {1 \over 4} \sum_{i=1}^2 \sum_{k=1}^2 \Big[ \, | \langle g_k | e_i \rangle |^4 + | \langle g_k | f_i \rangle |^4 \, \Big] \, .
\end{eqnarray}
Using Eqs.~(\ref{eq:basisStatesEveKMB}) and (\ref{eq:basisStatesEveKMB3}), this expression simplifies to 
\begin{eqnarray} \label{eq:P_ITER_KMB2}
P_{\rm ITER} &=& | \langle g_1 | e_1 \rangle |^2 + | \langle g_1 | f_1 \rangle |^2 -  | \langle g_1 | e_1 \rangle |^4 \nonumber \\
&& - | \langle g_1 | f_1 \rangle |^4 \, .
\end{eqnarray}
This expression can be calculated relatively easily for all possible bases $e$, $f$, and $g$, i.e.~for all possible values of $\theta_1$, $\theta_3$, and $\phi_3$.

\subsection{Quantum bit errors}

In addition to calculating the transmission error rate in terms of the index, Alice and Bob can calculate the transmission error rate in terms of the quantum bit. The Quantum Bit Error Rate (QBER) is calculated by selecting a certain number of control bits from the obtained key sequence and comparing them openly. For the KMB09 protocol with two-dimensional photon states, the probability $P_{\rm QBER}$ for getting a wrong bit equals \cite{KMB2009}
\begin{eqnarray}\label{eq:QBER_KMB}
P_{\rm QBER} &=& {4 - \sum_{i=1}^2 \sum_{k=1}^2 \Big[ \, | \langle e_i | g_k \rangle |^4 + | \langle f_i | g_k \rangle |^4 \, \Big] \over 8 - \sum_{i=1}^2 \sum_{k=1}^2 \Big[ \, | \langle e_i | g_k \rangle |^2 + | \langle f_i | g_k \rangle |^2 \, \Big]^2} \, . \nonumber \\
\end{eqnarray} 
Using again Eqs.~(\ref{eq:basisStatesEveKMB}) and (\ref{eq:basisStatesEveKMB3}), the above equation simplifies to 
\begin{eqnarray}\label{eq:QBER_KMB_Simple}
\lefteqn{ P_{\rm QBER} = } \nonumber \\ && { \, | \langle e_1 | g_1 \rangle |^2 + | \langle f_1 | g_1 \rangle |^2 - | \langle e_1 | g_1 \rangle |^4 - | \langle f_1 | g_1 \rangle |^4 \, \over \, 2 ( | \langle e_1 | g_1 \rangle |^2 + | \langle f_1 | g_1 \rangle |^2 ) - (| \langle e_1 | g_1 \rangle |^2 + | \langle f_1 | g_1 \rangle |^2)^2  \, } \, . \nonumber \\
\end{eqnarray} 
Using the concrete vectors in Eq.~(\ref{eq:general_efh_KMB}) we can now calculate the QBER for all possible scenarios.

\subsection{Efficiency with and without eavesdropping}

In the KMB09 protocol, Alice and Bob successfully obtain a key bit when both use different bases and when the index of the state measured by Bob is different from the index of Alice's state. The rate of transmitted key bits per transmitted photon hence equals \cite{KMB2009}
\begin{eqnarray}\label{eq:efficiency_KMB}
\eta &=& {1 \over 2} \cdot  {1 \over 4} \sum_{i=1}^2 \sum_{j\neq i} \Big[ \, | \langle e_{i} | f_{j} \rangle |^2 + | \langle f_{i} | e_{j} \rangle |^2 \, \Big] \, ~~~~
\end{eqnarray}
in the absence of any eavesdropping. This equation takes into account that Alice chooses randomly between 4 different states and that there is a probability of ${1 \over 2}$ for Bob to select one of the two bases $e$ and $f$. Under ideal conditions, no key bit is obtained whenever Alice and Bob use the same bases. Renaming some of the indices and using again the relations in Eqs.~(\ref{eq:basisStatesEveKMB}) and (\ref{eq:basisStatesEveKMB3}), Eq.~(\ref{eq:efficiency_KMB}) simplifies to
\begin{eqnarray}\label{eq:efficiencyKMB_efh1}
\eta &=& {1 \over 2} - {1 \over 2} \, | \langle e_1 | f_1 \rangle |^2 \, .
\end{eqnarray}
For the concrete basis vectors in Eq.~(\ref{eq:general_efh_KMB}), we hence find
\begin{eqnarray}\label{eq:efficiencyKMB_efh1}
\eta &=& {1 \over 2} \sin^2 \left( {1 \over 2}\theta_1 \right) \, .
\end{eqnarray}
For completeness, we also calculate the bit transmission rate $\eta_{\rm Evan}$ in the case of an intercept-resend eavesdropping attack. This efficiency equals the probability that the index $j$ of Bob's state is different from the index $i$ of the state send by Alice which is given by \cite{KMB2009}
\begin{eqnarray}\label{eq:efficiencyevan_KMB}
\eta_{\rm Evan} &=& {1 \over 8} \sum_{i=1}^2 \sum_{k=1}^2 \sum_{j\neq i} \Big[ \, | \langle e_{i} | g_{k} \rangle |^2 . | \langle g_{k} | e_{j} \rangle |^2 \nonumber \\
&& + | \langle f_{i} | g_{k} \rangle |^2 . | \langle g_{k} | f_{j} \rangle |^2 + | \langle f_{i} | g_{k} \rangle |^2 . | \langle g_{k} | e_{j} \rangle |^2
\nonumber \\
&& + | \langle e_{i} | g_{k} \rangle |^2 . | \langle g_{k} | f_{j} \rangle |^2 \, \Big] \, . ~~~~
\end{eqnarray}
Using again Eqs.~(\ref{eq:basisStatesEveKMB}) and (\ref{eq:basisStatesEveKMB3}) one can show that this equation is equivalent to
\begin{eqnarray}\label{eq:efficiencyevan2_KMB}
\eta_{\rm Evan} &=& 1 - {1 \over 8} \sum_{i=1}^2 \sum_{k=1}^2 \Big[ \, | \langle e_{i} | g_{k} \rangle |^2 + | \langle f_{i} | g_{k} \rangle |^2 \, \Big]^2 \, . ~~~~
\end{eqnarray}
Renaming indices and using again Eqs.~(\ref{eq:basisStatesEveKMB}) and (\ref{eq:basisStatesEveKMB3}) finally yields  
\begin{eqnarray}\label{eq:efficiencyevan3_KMB}
\eta_{\rm Evan} &=& | \langle e_1 | g_1 \rangle |^2 + | \langle f_1 | g_1 \rangle |^2 \nonumber \\
&& - {1 \over 2} \, \Big[ \, | \langle e_1 | g_1 \rangle |^2 + | \langle f_1 | g_1 \rangle |^2  \, \Big]^2 \, . ~~~
\end{eqnarray}
This bit transmission rate is in general much higher than the efficiency $\eta$ in Eq.~(\ref{eq:efficiencyKMB_efh1}). 

\subsection{Interplay between the ITER and the QBER} \label{complex}

\begin{figure}[b]
	\centering
	(a)
  	\includegraphics[width = 3.5 in]{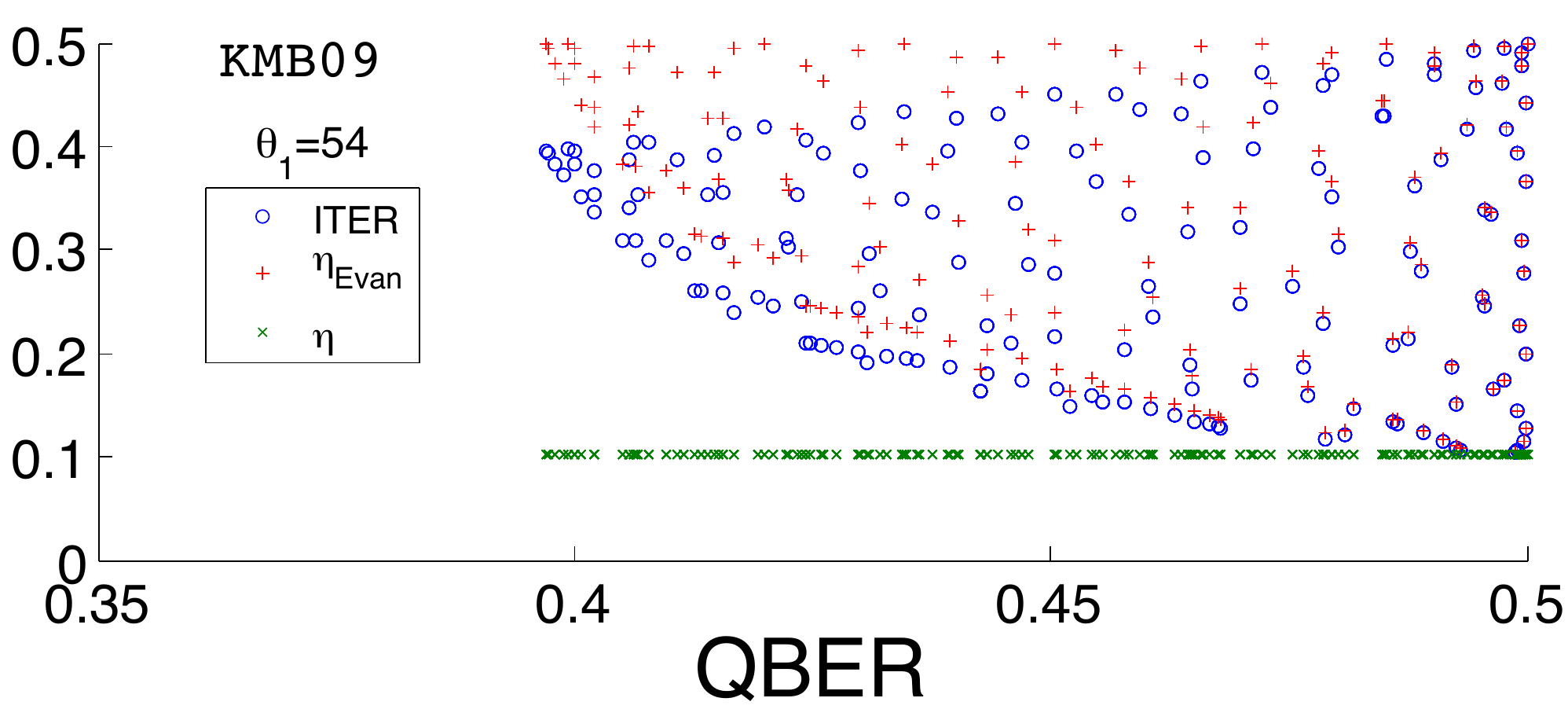}
	(b)
	\includegraphics[width = 3.5 in]{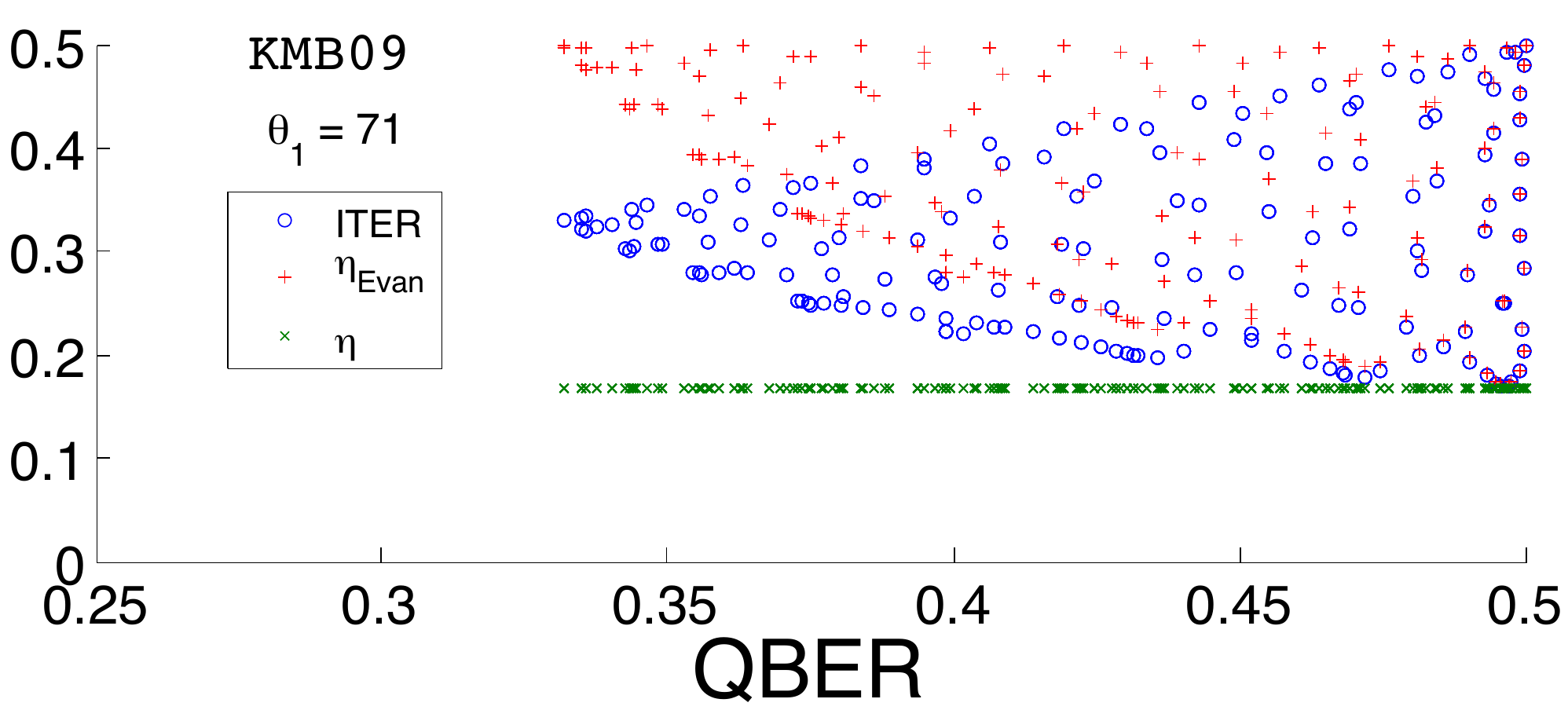}
	(c)
	\includegraphics[width = 3.5 in]{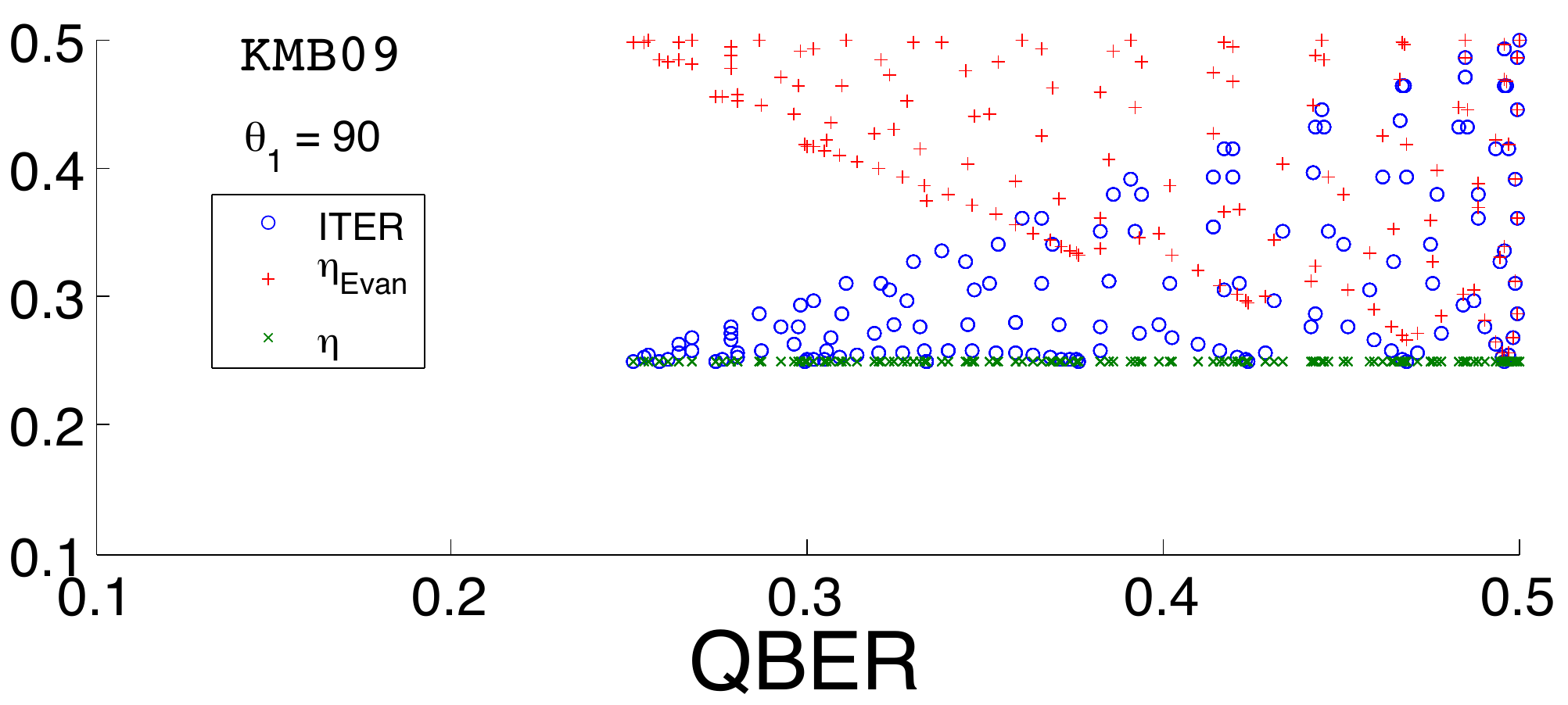}
\caption{The ITER, $\eta$ and $\eta_{\rm Evan}$ as a function of the corresponding quantum bit error rate for (a) $\theta_{1} = 54^\circ$, (b) $\theta_{1} = 71^\circ$, and (c) $\theta_{1} = 90^\circ$ for all possible choices of Evan's measurement basis. This means, the above figures are the result of a numerical simulation which varies both angles $\theta_3$ and $\phi_3$ between $0$ and $2 \pi$.}
	\label{fig:EvanOpt_KMB09}
\end{figure}

As mentioned in Section \ref{intro}, one feature of KMB09 is that an eavesdropper introduces not only one but {\em two} different types of errors into the key bit transmission. The relationship between the QBER, the ITER and the key bit transmission efficiency $\eta_{\rm Evan}$ can be used by Alice and Bob to detect eavesdropping much more efficiently. Evan only remains unnoticed, if these variables change, as if their changes were caused by system errors and not by eavesdropping. When there are very strong correlations between error rates and efficiencies, finding an eavesdropping strategy which fulfills this condition becomes a very hard, maybe even an impossible task for Evan to complete. 

We therefore now have a closer look at the relation between error rates and key bit transmission efficiencies for KMB09. In the following we consider three different choices of Alice and Bob's angle $\theta_1$ and assume that $\theta_{1}$ equals either $54^\circ$, $71^\circ$, or $90^\circ$. Moreover, we consider all possible choices of Evan's measurement basis $g$ by varying both angles $\theta_3$ and $\phi_3$ in Eq.~(\ref{eq:general_efh_KMB}) between $0$ and $2 \pi$. For each case, we calculate the QBER, the ITER, and the bit transmission efficiencies $\eta$ and $\eta_{\rm Evan}$. As illustrated in Fig.~\ref{fig:EvanOpt_KMB09}, the QBER in case of eavesdropping is in all three cases relatively high. It equals $40 \, \%$, $33 \, \%$, and $25 \, \%$, respectively. We also see that $\eta$ is relatively high in all three cases. It equals $12 \, \%$, $17 \, \%$, and $25 \, \%$, respectively. 

Moreover we notice that Evan has only limited freedom to choose certain values for the QBER, the ITER, and the key bit transmission efficiency $\eta_{\rm Evan}$, as long as the QBER is relatively close to its theoretical minimum. For example, for $\theta_{1} = 54^\circ$, we observe that minimising the QBER yields $P_{\rm ITER} = 0.4$ and $\eta_{\rm Evan} = 0.5$. Minimising the ITER is possible but not without increasing the QBER further. However, for relatively large QBER's, Evan has in general quite a lot of flexibility to choose his measurement basis $g$ such that the ITER, the QBER, and $\eta_{\rm Evan}$ assume certain values by carefully adjusting $\theta_3$ and $\phi_3$. In other words, Evan is able to make eavesdropping errors look like system errors when it matters most. To overcome this problem, we now have a look at a variation of the KMB09 protocol. 

\section{A variation of the KMB09 protocol} \label{More}

There are many possible generalisations of the above described protocol (cf.~eg.~Refs.~\cite{KMB2009} and \cite{BS2009}). In the following, we present a variation which replaces the two bases $e$ and $f$ of Alice and Bob by three bases $e$, $f$, and $h$. Again we restrict ourselves to two-dimensional photon states. As we shall see below, the proposed variation of KMB09 yields QKD protocols with strong correlations between the ITER and the QBER. This means, errors caused by intercept-resend eavesdropping have a distinct signature. 

\subsection{Basic premise}

In the following we denote the bases used by Alice and Bob $e$, $f$, and $h$ such that
\begin{eqnarray}\label{eq:basisStatesGeneral}
	e &\equiv & \left\{ \left| e_{i}  \right\rangle: i=1,2 \right\} \, , \nonumber \\ 
	f &\equiv & \left\{ \left| f_{i}  \right\rangle: i=1,2 \right\} \, , \nonumber \\ 
	h &\equiv & \left\{ \left| h_{i} \right \rangle: i=1,2 \right\} \, .
\end{eqnarray}
Suppose these bases encode 0=`$\mathtt{00}$', 1=`$\mathtt{01}$', and 2=`$\mathtt{10}$', respectively. Proceeding as above and using a table similar to Table \ref{tab:states_interpretation1}, Bob can then no longer decode Alice's key bits even when both obtain states with a different index. Suppose Alice sends a photon in $|e_i \rangle$ and Bob measures $|h_j \rangle$ with $i \neq j$. Then Bob knows that the incoming photon has not been prepared in the \textit{h} basis but he still does not know whether Alice sends a 0 or a 1.

\begin{table}[b]
\begin{center}
\begin{tabular}{@{}|c|c|c|c|}
\hline
& $S_1$ & $S_2$ & $S_3$ \\
\hline
  $e$ & 0 & 0 & --  \\
\hline
   $f$  & 1 & -- & 0 \\
\hline
   $h$  & -- & 1 & 1 \\
\hline 
\end{tabular}
\end{center}
  	\caption{Meaning of Alice's basis in which she prepares the photon depending on which set has been announced by Bob.}
  \label{tab:encoding23}
    \end{table}

A solution to this problem which does not result in a significant drop of the key bit transmission efficiency is that Alice and Bob form three pairs by pairing each basis with every other basis. In the following we denote these sets by $S_1$, $S_2$, and $S_3$ and define
\begin{eqnarray} \label{Si}
S_1 \equiv \left\{e,f\right\} \, , ~~ S_2 \equiv \left\{e,h\right\} \, , ~~ S_3 \equiv \left\{f,h\right\} \, . 
\end{eqnarray}
Suppose Alice sends a photon prepared in $|e_i \rangle$ which is measured by Bob in $|h_j \rangle$ with $i \neq j$. Bob now randomly chooses one set which includes his measurement basis and announces it publicly. If he announces for example $ S_2 $, then Alice knows Bob's measurement basis which allows them to obtain a shared-secret key. If Bob announces $ S_3 $, then Alice does not know Bob's measurement basis. In this case, Alice and Bob cannot deduce a key bit although both of their states have a different index. 

In order to avoid that Evan can guess key bits with a probability that is higher than the probability for correct random guessing, Alice and Bob should employ the analogy of their situation to the KMB09 protocol and restrict themselves to a two letter alphabet. They should use a different encoding of the transmitted key bit, depending on which set of basis pairs has been announced by Bob. One possible encoding is shown in Table~\ref{tab:encoding23}: here $f$ encodes either a 0 or a 1, while $e$ always encodes 0, and $h$ always encodes 1. Most importantly, each set of bases $S_k$ is equally likely to encode a 0 and to encode a 1. In this way, the eavesdropper obtains no information when Bob announces his chosen set. How this new encoding affects Bob's interpretation of his measurement outcomes is explained in Table~\ref{tab:encoding2}. 

Before stating the concrete protocol let us summarise the three types of classical information which Alice and Bob have to reveal with every successful photon transmission:
\begin{enumerate}
	\item [a]Firstly, Alice must announce the index of each transmitted state after sending the photon to Bob. 
	\item [b]Secondly, Bob must announce a set $S_k$, whenever Bob's measurement index is different from Alice's index.
        \item [c]Finally, Alice must announce whether Bob's set $S_k$ contains her basis or not. 
\end{enumerate}
Bob chooses his set $ S_k $ by randomly choosing between two bases pairs which contain his measurement basis.  A key bit is obtained, if the set chosen by Bob contains both the basis which the photon has been prepared in and the basis which the photon has been measured in. 

\begin{table}[t]
    \begin{center}
				\begin{tabular}{|c|c|c|c|c|c|c|}
					\hline
					 & \multicolumn{6}{|c|}{Bob's bases} \\ \cline{2-7}
					Alice's  & \multicolumn{2}{|c|}{$S_{1}$}  & \multicolumn{2}{|c|}{$S_{2}$} & \multicolumn{2}{|c|}{$S_{3}$} \\ \cline{2-7}
					bases    & $e$      &	$f$        & $e$      &	$h$      & $f$     & $h$ \\
					\hline
				  	$e$ meaning 0 & error &	$0$        & error &	$0$      & -     & - \\
					$f$ meaning 0 or 1 & $1$      &	error   & -     &	-     & error &	$0$ \\
					$h$ meaning 1 & -      &	-        & $1$      &	error & $1$      & error  \\
   					\hline
  				\end{tabular}\centering
    \end{center}
\caption{Interpretation of Bob's measurement outcomes for the case, where the indices of Alice and Bob's states are different and where both their bases are included in the set announced by Bob. Evan cannot guess which bit is transmitted, since each set is equally likely to encode a 0 or a 1.}
  \label{tab:encoding2}
\end{table}

\subsection{The proposed protocol} \label{proto}

We now assume that Alice and Bob already agreed on a concrete set of bases $\textbf{B} =  \left\{e, f, h \right\}$ for the preparation and the measurement of their photon states. Moreover, we assume that they introduced the basis pairs $S_1$, $S_2$, and $S_3$ in Eq.~(\ref{Si}) and that they agreed to use the interpretation of information suggested in Tables \ref{tab:encoding23} and \ref{tab:encoding2}. Then their QKD protocol works as follows:
\begin{enumerate}
	\item Alice generates a random sequence of symbols $x \in \{e,f,h \}$ and assigns a random index $ i \in \{1,2\}$ to every one of them. 
	\item Alice then uses this sequence and the corresponding indices, prepares a sequence of photons in the respective states $|x_i \rangle$, and sends them to Bob.
	\item Bob measures the state of every incoming photon, thereby randomly switching his measurement basis among the three bases in \textbf{B}.
	\item Afterwards, Alice announces her random sequence of indices $i$ publicly.
	\item Bob announces which photons he found in a state $|y_j \rangle$ with an index $j$ which is different from the $i$ of the corresponding state $|x_i \rangle$. For each of these photons, he announces a set $ S_k $ which contains his respective measurement basis. 
	\item Alice announces which ones of these sets contain the basis in which she prepared the respective photon. Now Alice and Bob both know which photons transmit one key bit and interpret the corresponding photon states accordingly. 
	\item Finally,  Alice and Bob openly compare some quantum bits and some indices of transmitted states. This allows them to calculate their QBER, their ITER, and their key bit transmission efficiency. If they detect an eavesdropper, the generated cryptographic key is not secure.
\end{enumerate}

\subsection{Parametrisation of states}

Eq.~(\ref{eq:general_efh_KMB}) describes a way to parametrize the basis states of $e$, $f$, and $g$ by three angles $\theta_1$, $\theta_3$, and $\phi_3$. Since the QKD protocol discussed in this section requires an additional basis $h$, we moreover define
\begin{eqnarray} \label{eq:h} 
&& \hspace*{-0.7cm} \left| h_{1} \right\rangle = \left( \begin{array}{r}  \cos \left( \frac{1}{2} \theta_2 \right) \\   {\rm e}^{{\rm i}\phi_2} \sin \left( \frac{1}{2} \theta_2 \right) \end{array} \right) \, , ~~
\left| h_{2} \right\rangle  = \left( \begin{array}{r}  \sin \left( \frac{1}{2} \theta_2 \right) \\ - {\rm e}^{{\rm i}\phi_2}\cos \left( \frac{1}{2} \theta_2 \right) \end{array} \right) \, , \nonumber \\
\end{eqnarray}
where $\theta_2$ and $\phi_2$ are real numbers ranging from $0$ to $2\pi$.

\subsection{Index transmission errors}

An index transmission error occurs when a photon prepared in a state with index $i$ is measured at Bob's end in the same basis but found in a state with an index $j \neq i$. To calculate the corresponding error rate we consider only the cases, where Alice and Bob use the same basis. Taking into account that the probability for choosing one of the six states available to Alice equals ${1 \over 6}$, we find that the probability for an index transmission error equals
\begin{eqnarray}\label{eq:P_ITER1}
P_{\rm ITER} &=& {1 \over{6}} \sum_{i=1}^2 \sum_{k=1}^2 \sum_{j \neq i} \Big[ \, 
| \langle e_{i} | g_k \rangle |^2 | \langle g_k | e_{j} \rangle |^2 + \nonumber \\ 
&&  \hspace*{-1cm} + | \langle f_{i} | g_k \rangle |^2 | \langle g_k | f_{j} \rangle |^2 + 
| \langle h_{i} | g_k \rangle |^2 | \langle g_k | h_{j} \rangle |^2 \, \Big] \, . ~~~
\end{eqnarray}
This equation adds the probabilities of all events which might result in the transmission of a wrong index, like the case where Alice sends a photon in $|e_1 \rangle$, Evan measures $|g_1 \rangle$, and Bob finally finds the photon forwarded to him by Evan in $|e_2 \rangle$. Using Eq.~(\ref{eq:basisStatesEveKMB}) and the fact that the states of all the bases are of unit length, Eq.~(\ref{eq:P_ITER1}) simplifies to
\begin{eqnarray}\label{eq:P_ITER3}
P_{\rm ITER} &=& 1 - {1 \over{6}} \sum_{i=1}^2 \sum_{k=1}^2 \Big[ \, | \langle g_k | e_{i} \rangle |^4 + | \langle g_k | f_{i} \rangle |^4 \, \nonumber \\
& & + | \langle g_k | h_{i} \rangle |^4 \, \Big] \,. 
\end{eqnarray}
As expected, the ITER is one minus the probability that no index transmission error occurs, when Alice and Bob measure in the same bases. Simplifying the ITER further using Eq.~(\ref{eq:basisStatesEveKMB3}) and analog relations for the $f$ and the $h$ basis, we find that
\begin{eqnarray}\label{eq:P_ITER4}
P_{\rm ITER} &=& {2 \over 3} \, \Big[ \, | \langle g_1 | e_1 \rangle |^2 + | \langle g_1 | f_1 \rangle |^2 + | \langle g_1 | h_1 \rangle |^2  \nonumber \\
&& - | \langle g_1 | e_1 \rangle |^4 - | \langle g_1 | f_1 \rangle |^4 - | \langle g_1 | h_1 \rangle |^4 \, \Big] ~~~
\end{eqnarray}
which is easier to evaluate than Eq.~(\ref{eq:P_ITER3}).

\subsection{Quantum bit errors}

The QBER is the ratio of the number of erroneous bits to the total number of detected bits. This means, to determine their QBER, Alice and Bob need to select some of the transmitted bits as test bits and compare their values publicly. A closer look at the above protocol shows that the QBER is closely related to the ITER. The reason is that a quantum bit error occurs whenever Alice and Bob obtain a different index, although both use the same basis (c.f.~Table~\ref{tab:encoding2}). When both use the same bases, it does not matter which set $S_1$, $S_2$, or $S_3$ Bob chooses, since Alice's basis is automatically included in his set and a key bit is obtained. Taking this and the fact that Alice and Bob both choose the same bases with probability ${1 \over 3}$ into account, we find that
\begin{eqnarray}\label{eq:QBER1}
P_{\rm QBER} &=& {P_{\rm ITER} \over 3 P_{\rm QB}} \, ,
\end{eqnarray}
where $P_{\rm QB} $ is the probability of Alice and Bob obtaining a key bit for each transmitted photon. 

To calculate $P_{\rm QB} $, we notice that whenever Alice and Bob use a different basis, there is a probability of ${1 \over 2}$ that Bob chooses a set $S_1$, $S_2$, or $S_3$ which does not contain Alice's basis. This means, even when Alice and Bob measure a different basis and obtain a different index, no key bit is obtained in half of the cases. To calculate $P_{\rm QB} $ we now sum over all these cases and the cases where Alice and Bob contain a wrong bit and find that 
\begin{eqnarray}\label{eq:P_DI1}
P_{\rm QB} &=& {1 \over 3} P_{\rm ITER} + {1 \over 2} \cdot {1 \over 3} \cdot {1 \over 6} \sum_{i=1}^{2} \sum_{k=1}^{2} \sum_{j\neq i} \nonumber \\
&& \hspace*{-0.5cm} \Big[ \,  | \langle e_{i} | g_k \rangle |^2  | \langle g_k | f_{j} \rangle |^2 + | \langle e_{i} | g_k \rangle |^2  | \langle g_k | h_{j} \rangle |^2  \nonumber \\
&& \hspace*{-0.5cm}  + | \langle f_{i} | g_k \rangle |^2  | \langle g_k | e_{j} \rangle |^2 + | \langle f_{i} | g_k \rangle |^2  | \langle g_k | h_{j} \rangle |^2  \nonumber \\
&& \hspace*{-0.5cm} + | \langle h_{i} | g_k \rangle |^2  | \langle g_k | e_{j} \rangle |^2 + | \langle h_{i} | g_k \rangle |^2  | \langle g_k | f_{j} \rangle |^2 \, \Big] \, . ~~~~ 
\end{eqnarray}
The factor ${1 \over 3}$ is the probability of Bob choosing one out of the three possible bases $e$, $f$ and $h$, while the factor ${1 \over 6}$ takes into account that Alice prepares her photons randomly in one of six possible states.

Renaming indices in some of the terms, we see that every term in Eq.~(\ref{eq:P_DI1}) occurs twice. The quantum bit transmission probability $P_{\rm QB}$ hence simplifies to
\begin{eqnarray}\label{eq:P_DI2}
P_{\rm QB} &=& {1 \over 3} P_{\rm ITER} + {1 \over 18} \sum_{i=1}^{2} \sum_{k=1}^{2} \sum_{j\neq i} \Big[ \,  | \langle e_{i} | g_k \rangle |^2  | \langle g_k | f_{j} \rangle |^2 \nonumber \\
&&  \hspace*{-0.3cm} + | \langle e_{i} | g_k \rangle |^2  | \langle g_k | h_{j} \rangle |^2 + | \langle h_{i} | g_k \rangle |^2  | \langle g_k | f_{j} \rangle |^2 \, \Big] \, . ~~ 
\end{eqnarray}
Using relations like the ones in Eqs.~(\ref{eq:basisStatesEveKMB}) and (\ref{eq:basisStatesEveKMB3}), one can moreover show that this probability can also be written as 
\begin{eqnarray}\label{eq:P_DI3}
P_{\rm QB} &=& {4 \over 9} \, \Big[ \, | \langle g_1 | e_1 \rangle |^2 + | \langle g_1 | f_1 \rangle |^2 + | \langle g_1 | h_1 \rangle |^2 \Big] \nonumber \\
&& - {2 \over 9} \, \Big[ \, | \langle g_1 | e_1 \rangle |^4 + | \langle g_1 | f_1 \rangle |^4 + | \langle g_1 | h_1 \rangle |^4 
\nonumber \\
&& + | \langle g_1 | e_1 \rangle |^2 | \langle g_1 | f_1 \rangle |^2 + | \langle g_1 | e_1 \rangle |^2 | \langle g_1 | h_1 \rangle |^2 \nonumber \\
&& + | \langle g_1 | f_1 \rangle |^2 | \langle g_1 | h_1 \rangle |^2 \, \Big] ~~~
\end{eqnarray}
which is easier to calculate than Eq.~(\ref{eq:P_DI2}). 

\subsection{Efficiency with and without eavesdropping}

As in KMB09, the efficiency of the key bit transmission depends on whether Evan is present or not. For example, the key bit transmission rate $\eta_{\rm Evan}$ is the probability of Alice and Bob obtaining a key bit, when Evan listens into the conversation between Alice and Bob. This immediately implies 
\begin{eqnarray}\label{eq:efficiency_eve_efh}
\eta_{\rm Evan} &=& P_{\rm QB} 
\end{eqnarray} 
with $P_{\rm QB}$ as in Eq.~(\ref{eq:P_DI3}). In order to calculate the key transmission efficiency $\eta$ in the absence of eavesdropping, we have to take into account that Alice chooses randomly between six states, that there is a probability of ${1 \over 2}$ for Bob to announce a set which contains Alice's basis, and that there is a probability of ${1 \over 3}$ for Bob to select one of the three basis $e$, $f$, or $h$. Moreover, we notice again that no key bit is obtained whenever Alice and Bob use the same bases. As a result, we find that 
\begin{eqnarray}\label{eq:efficiency_efh}
\eta &=& {1 \over 2} \cdot {1 \over 3} \cdot {1 \over 6} \sum_{i=1}^{2} \sum_{j\neq i} \Big[ \, | \langle e_{i} | f_{j} \rangle |^2 + | \langle e_{i} | h_{j} \rangle |^2 \nonumber \\
&&  \hspace*{-0.7cm}  + | \langle f_{i} | e_{j} \rangle |^2 + | \langle f_{i} | h_{j} \rangle |^2 + | \langle h_{i} | e_{j} \rangle |^2 +  | \langle h_{i} | f_{j} \rangle |^2 \, \Big] \, . ~~~~
\end{eqnarray}
Renaming again some of the indices and using again relations like the ones in Eqs.~(\ref{eq:basisStatesEveKMB}) and (\ref{eq:basisStatesEveKMB3}), $\eta$ simplifies to
\begin{eqnarray}\label{eq:efficiency_efh1}
\eta &=& {1 \over 3} - {1 \over 9} \Big[ \, | \langle e_1 | f_1 \rangle |^2 + | \langle e_1 | h_1 \rangle |^2  + | \langle f_1 | h_1 \rangle |^2 \, \Big] \, . ~~~~
\end{eqnarray}
Using Eqs.~(\ref{eq:general_efh_KMB}) and (\ref{eq:h}), this efficiency $\eta $ becomes
\begin{eqnarray}\label{eq:efficiency_efh1}
\eta &=& {1 \over 6} - {1 \over 18} \left[ \cos \theta_1 + \cos \theta_2 + \cos \theta_1 \cos \theta_2 \right. \nonumber \\
&& \left. + \cos \phi_2 \, \sin \theta_1 \sin \theta_2 \right] \, .
\end{eqnarray}
Since this expression is not overly complicated, we add it here for convenience.   

\subsection{Interplay between the ITER and the QBER}

\begin{figure}[t]
	\centering
	(a)
  	\includegraphics[width = 3.5 in]{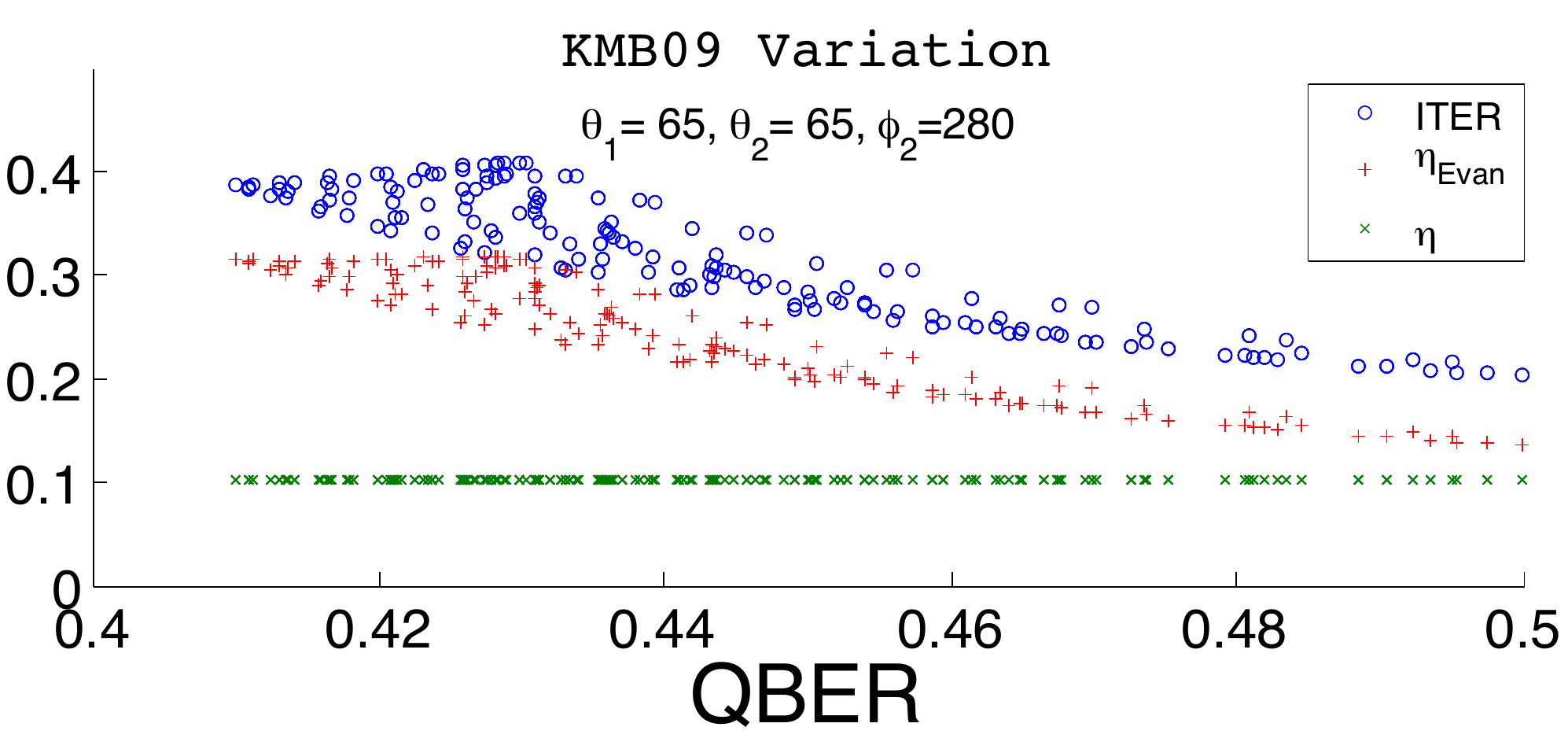}
	(b)
	\includegraphics[width = 3.5 in]{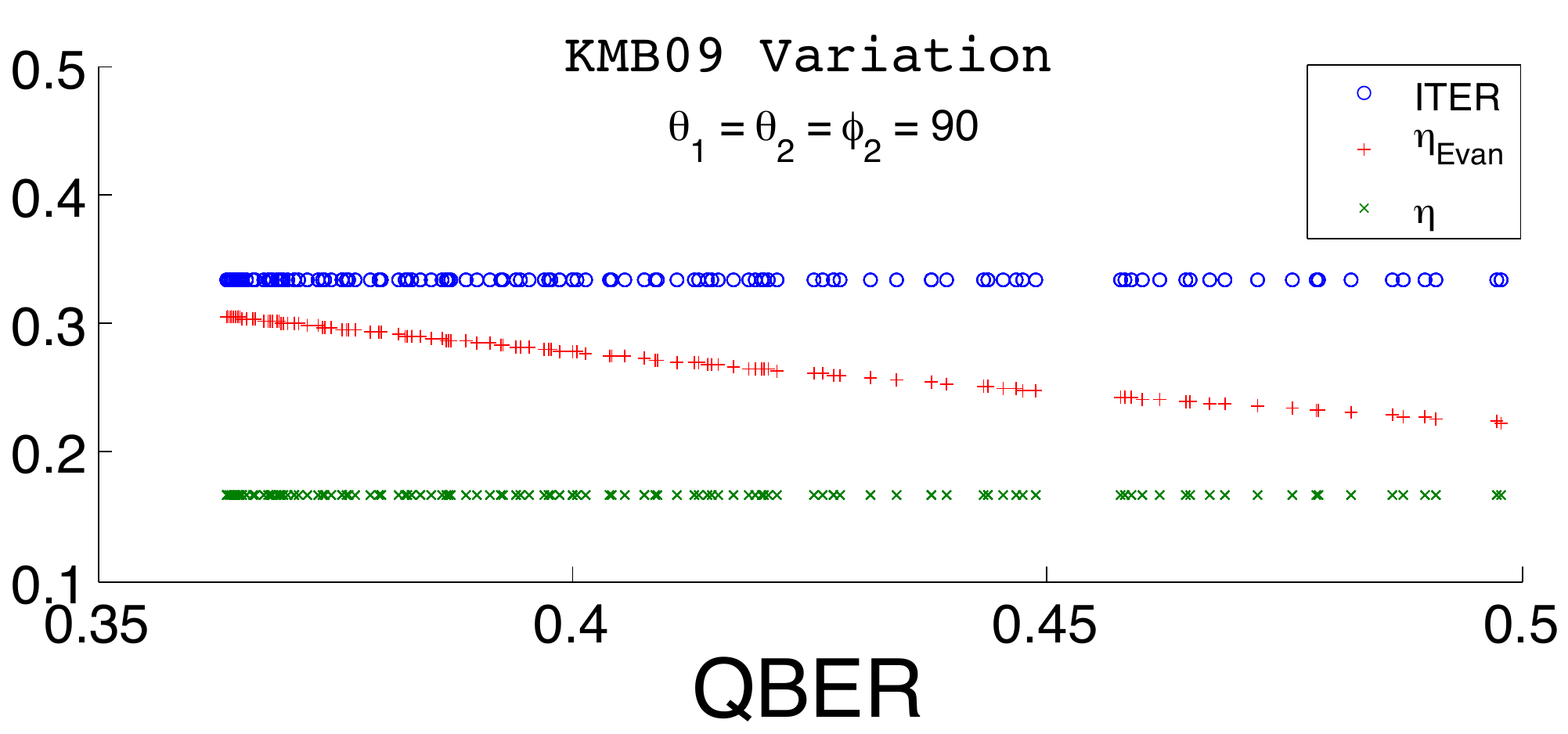}
	(c)
	\includegraphics[width = 3.5 in]{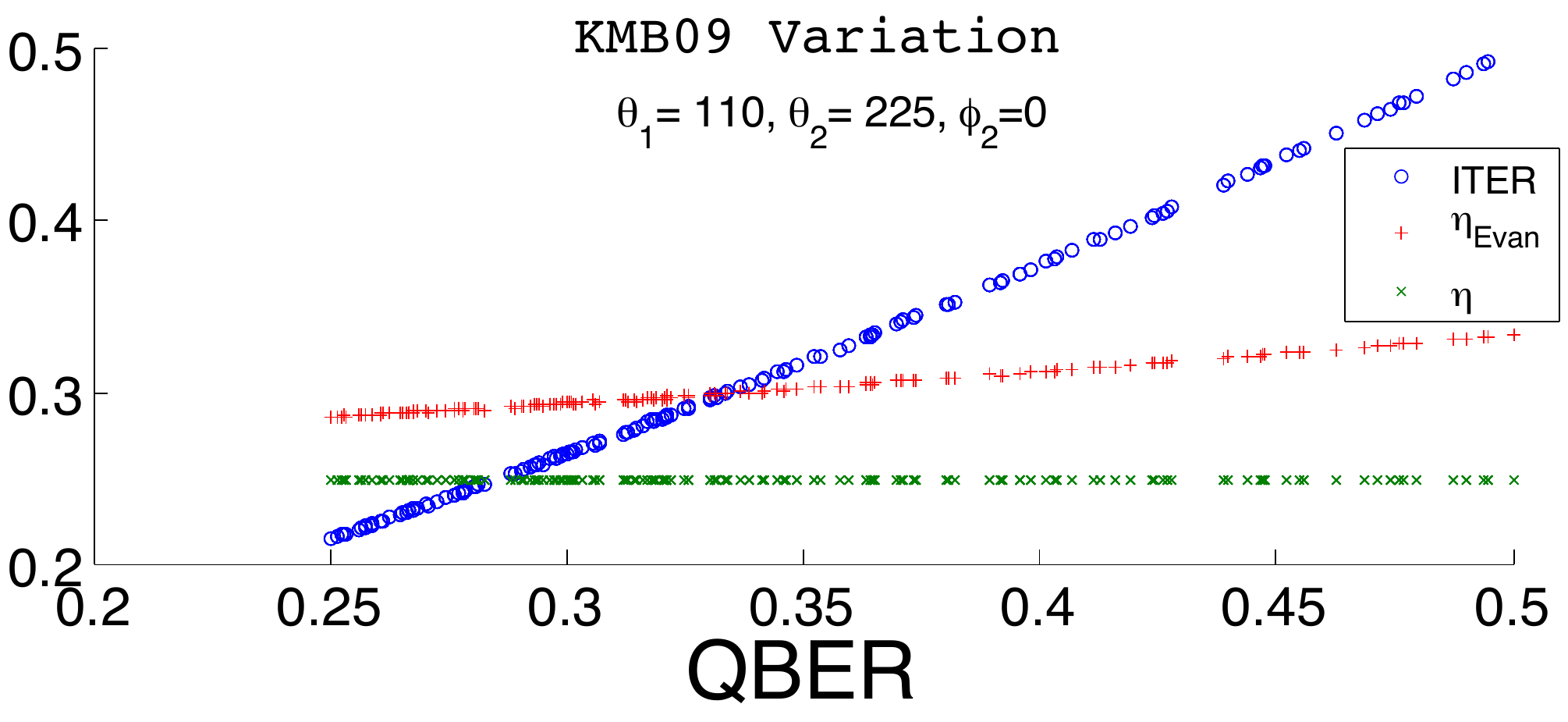}
  \caption{The ITER, $\eta$ and $\eta_{\rm Evan}$ as a function of the corresponding quantum bit error rate for (a) $\theta_{1} = 65^\circ, \theta_{2} = 65^\circ, \phi_{2} = 280^\circ$, (b) $\theta_{1} = 90^\circ, \theta_{2} = 90^\circ, \phi_{2} = 90^\circ$, and (c) $\theta_{1} = 110^\circ, \theta_{2} = 225^\circ, \phi_{2} = 0^\circ$ for all possible choices of Evan's measurement basis. Like Fig.~\ref{fig:EvanOpt_KMB09}, this figure is the result of a numerical simulation which varies both angles $\theta_3$ and $\phi_3$ between $0$ and $2 \pi$.}
	\label{fig:EvanOpt_KMBVar}
\end{figure}

The main difference between the QKD protocol which we discuss in this section and its original version is that using three bases gives Alice and Bob more flexibility. They can now choose their angles $\theta_1$, $\theta_2$, and $\phi_2$ such that there are very strong correlations between the QBER and the ITER in case of intercept-resent eavesdropping. Fig.~\ref{fig:EvanOpt_KMBVar} illustrates the interplay between error rates and efficiencies for three different choices of the bases $e$, $f$, and $h$.  Like Fig.~\ref{fig:EvanOpt_KMB09} before, this figure is the result of a numerical simulation which varies both angles $\theta_3$ and $\phi_3$ between $0$ and $2 \pi$. The main difference between Figs.~\ref{fig:EvanOpt_KMB09} and \ref{fig:EvanOpt_KMBVar} is that there is now a strong dependence of the ITER on the QBER. Figs.~\ref{fig:EvanOpt_KMBVar}(b) and (c) even show a linear dependence. This means, in these two cases, intercept-resend eavesdropping has a distinct signature which makes it much harder for an eavesdropper to hide his presence. If the system errors of the transmission line do not have the same signature, he cannot enter the scene unnoticed by simply replacing the transmission line between Alice and Bob with a less faulty one. In other words, eavesdropping errors can no longer be mimicked by system errors.

\section{Conclusions} \label{Conclusions}

This paper proposes a QKD protocol for which eavesdropping introduces not only one but two different types of errors into the key bit transmission with a distinct correlation between both of them. This is possible, since the protocol uses an encoding of information which is different from the encoding of information in BB84 \cite{BB84}. Instead of each basis encoding a whole alphabet, all the states of the same basis encode the same letter. As a result, eavesdropping introduces quantum bit  as well as index transmission errors. A quantum bit error occurs, when Alice and Bob both obtain different key bits. An index transmission error occurs when Alice and Bob use the same basis but their respective states nevertheless have different indices. Intercept-resend eavesdropping results, in general, in a significant change of both the Quantum Bit Error Rate (QBER) and the Index Transmission Error Rate (ITER). 

The main purpose of the proposed QKD protocol is to make it much harder for an eavesdropper to conceal his presence. The eavesdropper now not only has to minimise different types of errors. The above mentioned strong correlations between the ITER and the QBER introduce a distinct signature of eavesdropping. This signature allows Alice and Bob to detect the origin of their errors and to tolerate lower eavesdropping and higher system error rates without compromising their privacy. We no longer require that the error rates in case of eavesdropping unavoidably exceed the system error rate, since Evan can no longer disguise his presence as system errors. As a result, Alice and Bob should be able to communicate securely over much longer distances.

Finally, let us remark that, for concrete practical implementations of the proposed QKD scheme, Alice and Bob should choose their bases $e$, $f$, and $h$ such that their system errors (and not the eavesdropping errors) have a distinct signature. Changes of this signature can then herald Evan's presence, independent of his eavesdropping strategy. Evan only remains unnoticed, if the error rates and efficiencies which he introduces into the key bit transmission are the same as the ones caused by errors of the concrete experimental setup. Finding an eavesdropping strategy which fulfils this condition is in general a very hard task for Evan to complete. Our analysis suggests that it might be impossible for Evan to satisfy this condition. \\[0.5cm]

{\em Acknowledgment.} M. M. K. acknowledges funding from the NED University of Engineering \& Technology, Karachi, Pakistan.


\begin{thebibliography}{12}
\bibitem{grtz2002}
Gisin, N., Ribordy, G., Tittel, W., Zbinden, H.: {\em Reviews of Modern Physics.} 2002, {\em 74}, 145.

\bibitem{BB84}
Bennett, C.H., Brassard, G.: Quantum cryptography: Public key distribution and coin tossing. {\em Proceedings of IEEE International Conference on Computers, Systems, and Signal Processing}, Bangalore, India 1984 175.

\bibitem{B92}
Bennett, C. H.: Quantum Cryptography Using Any 2 Nonorthogonal States. Physical Review Letters {\em 68} 1992 3121.

\bibitem{DB1998}
Bru\ss, D.: Optimal Eavesdropping in Quantum Cryptography with Six States. Physical Review Letters {\em 81} 1998 3018.

\bibitem{BG1999}
Bechmann-Pasquinucci, H., Gisin, N.: Incoherent and coherent eavesdropping in the six-state protocol of quantum cryptography. Physical Review A {\em 59} 1999 4238.

\bibitem{Beige22002}
Beige, A., Englert, B., Kurtsiefer, C., Weinfurter, H.: Secure communication with single-photon two-qubit states. Journal of Physics A-Mathematical and General {\em 35} 2002 L407.

\bibitem{Beige12002}
Beige, A., Englert, B. G., Kurtsiefer, C., Weinfurter, H.: Secure communication with a publicly known key. Acta Physica Polonica A {\em 101} 2002 357.

\bibitem{KMB2009}
Khan, M. M., Murphy, M., Beige, A.: High  error rate quantum key distribution for long-distance communication. New Journal of Physics {\em 11} 2009 063043.

\bibitem{hugo2}
Scarani V., Acin A., Ribordy G., Gisin N.: Physical Review Letters {\bf 92} 2004 057901.

\bibitem{BS2009}
Brierley, S.: Quantum Key Distribution Highly Sensitive to Eavesdropping, unpublished 2009. arXiv:0910.2578.

\bibitem{BP2000}
Bechmann-Pasquinucci, H., Peres, A.: Quantum cryptography with 3-state systems. Physical Review Letters {\em 85} 2000 3313.

\bibitem{Knight} 
Gerry, C. C., Knight, P. L.: Introductory Quantum Optics (Cambridge: Cambridge University Press) 2005.

\end{thebibliography}
\end{document}